\PassOptionsToPackage{nameinlink}{cleveref}
\documentclass[a4paper,UKenglish,cleveref]{lipics-v2021}
\usepackage{nicefrac}
\usepackage{mathtools}
\usepackage{thm-restate}
\usepackage{xspace}
\usepackage{complexity}
\usepackage[normalem]{ulem}
\usepackage{siunitx}
\usepackage[table]{xcolor}

\graphicspath{{./figures/}}

\crefname{conjecture}{Conjecture}{Conjectures}

\usepackage{subcaption}
\usepackage{pifont}
\newcommand{\yes}{\cellcolor[RGB]{171,203,173}\ding{52}}
\newcommand{\no}{\cellcolor{red!20}\ding{56}}
\newcommand{\flipSeq}{\ensuremath{\mathcal{S}}\xspace}

\title{Counterexamples to the Happy Edge Conjecture in Plane Spanning Trees}
\titlerunning{Counterexamples to the happy edge conjecture in plane spanning trees}

\author{Oswin Aichholzer}{Institute of Algorithms and Theory, Graz University of Technology, Austria}{oswin.aichholzer@tugraz.at}{https://orcid.org/0000-0002-2364-0583}{}
\author{Joseph Dorfer}{Institute of Algorithms and Theory, Graz University of Technology, Austria}{joseph.dorfer@tugraz.at}{https://orcid.org/0009-0004-9276-7870}{Austrian Science Fund (FWF) 10.55776/DOC183.}
\author{Peter Kramer}{Department of Computer Science, TU Braunschweig, Germany}{kramer@ibr.cs.tu-bs.de}{https://orcid.org/0000-0001-9635-5890}{Supported by a fellowship of the German Academic Exchange Service~(DAAD) and the Deutsche Forschungsgemeinschaft (DFG, German Research Foundation), grant 530918134.}
\author{Christian Rieck}{Institute of Mathematics, University of Kassel, Germany}{christian.rieck@mathematik.uni-kassel.de}{https://orcid.org/0000-0003-0846-5163}{Funded by the Deutsche Forschungsgemeinschaft (DFG, German Research Foundation), grant  522790373.}
\author{Birgit Vogtenhuber}{Institute of Algorithms and Theory, Graz University of Technology, Austria}{birgit.vogtenhuber@tugraz.at}{https://orcid.org/0000-0002-7166-4467}{}

\authorrunning{O.~Aichholzer, J.~Dorfer, P.~Kramer, C.~Rieck, and B.~Vogtenhuber}
\Copyright{Oswin Aichholzer, Joseph Dorfer, Peter Kramer, Christian Rieck, and Birgit Vogtenhuber}



\hideLIPIcs
\nolinenumbers
\acknowledgements{This work was initiated at the Reconfiguration Week hosted at TU Graz in March~2025. We thank the organizers and all participants for fruitful discussions and the very good~atmosphere.}

\begin{document}
    \maketitle

    \begin{abstract}
        We study the problem of reconfiguring plane, straight-line spanning trees on point sets in the plane.
        A \emph{flip} is a reconfiguration step in which one edge of a tree is replaced with another, yielding again a plane spanning tree.
        For a given point set $P$, the \emph{flip graph} has a vertex for each plane spanning tree on~$P$, and an edge between any two trees that differ by a single flip.

        The folklore \emph{happy~edge conjecture} suggests that any edge shared by both the initial and the target tree (a \emph{happy} edge) is never flipped in a shortest flip sequence.
        This conjecture was computationally verified for up to 10 points in convex position.
        In this paper, we disprove this conjecture both on general point sets and on point sets in convex position by providing two counterexamples on~$16$ and~$30$ points, respectively.

        The counterexample for point sets in general position also holds for compatible flips, while the happy edge property holds for compatible flips in convex point sets as shown by Aichholzer, Dorfer, and Vogtenhuber~[GD 2025].
    \end{abstract}

    \section{Introduction}\label{sec:introduction}
    Reconfiguration problems ask whether one feasible state of a system can be transformed into another through a sequence of allowed moves.
    A classical example concerns the reconfiguration of triangulations of a convex point set.
    In this context, a \emph{flip} replaces the common diagonal of two adjacent triangles with the other diagonal of the convex quadrilateral they form.
    The resulting \emph{flip graph}, whose vertices are triangulations and whose edges correspond to flips, is the $1$-skeleton of the associahedron.
    These triangulations are in bijection with binary trees~\cite{SleatorTT86}, and each flip corresponds exactly to a rotation in the associated tree.
    Consequently, the minimum number of flips needed to transform any triangulation into any other equals the number of rotations needed to transform their corresponding binary trees.
    This provides an algorithmic motivation for studying shortest flip sequences: rotations are basic operations of binary trees, and having a fast method to compute shortest distances would immediately improve the efficiency of algorithms that are based on those data structures.

    While the diameter of the 1-skeleton of the associahedron is known exactly for all sizes~\cite{POURNIN201413}, determining the diameter of the flip graph of plane spanning trees remains open, even for points in convex position.
    Avis and Fukuda~\cite{AVIS199621} proved that, for any set of $n$ points in general position, this flip graph is connected and has diameter at most $2n-4$.
    The lower bound on the diameter of~$1.5 n-5$ by Hernando, Hurtado, Márquez, Mora, and Noy~\cite{HERNANDO199951} even holds for the special case of convex point sets.
    The currently best known lower bound is $(\nicefrac{11}{7}-o(1))n$, established by Bjerkevik, Dorfer, Kleist, Ueckerdt, and Vogtenhuber~\cite{bjerkevik2026flippingHardness}.
    For convex point sets, the upper bound was gradually improved~\cite{AichholzerBBDDKLLTU24,bousquet2025reconfiguration,bousquet2023notes}, with the currently best known bound being $\nicefrac{5n}{3}-3$ by Bjerkevik, Kleist, Ueckerdt, and Vogtenhuber~\cite{bjerkevik2025flipping}.

    \medskip
    The authors of~\cite{AichholzerBBDDKLLTU24} present their $2n-\log(n)$ bound as follows:

    \begin{quote}
        \emph{``...happy edges do not flip. An unhappy edge of $T_I$ flips once or twice, and if twice, it flips to the convex hull and then to an edge of $T_F$.''}
    \end{quote}

    Whether guided by structural evidence or a little wishful thinking, it is tempting to hope that shortest flip sequences also admit such a simple structure.
    In fact, the following conjecture, which captures exactly this property, is widely considered folklore.

    \begin{conjecture}[Happy edge conjecture]\label{conj:happy}
        Let $T_I$ and $T_F$ be two plane spanning trees, and let $H=T_I\cap T_F$ be the set of all happy edges.
        Then there exists a shortest flip sequence~$T_I=T_0,T_1,\ldots,T_{k-1},T_k=T_F$ such that $H\subset T_i$ for every $i\in[0,k]$.
    \end{conjecture}

    In other words, the happy edge conjecture postulates that any two plane spanning trees on a point set admit a shortest flip sequence that preserves every edge common to the initial and target trees.
    This conjecture has been a subject of significant interest for multiple decades, with several works providing improved bounds on the flip graph diameter by giving constructive arguments that satisfy this property~\cite{AichholzerBBDDKLLTU24, bjerkevik2025flipping,bousquet2025reconfiguration}.
    These constructions provided evidence that shortest sequences might also have this structure.
    Recent work has further attempted to strengthen the original conjecture and obtain additional structural properties.
    The \emph{parking edge conjecture}~\cite{AichholzerBBDDKLLTU24} intuitively states that shortest flip sequences on points in convex position use only edges in the union of trees and on their convex hull.
    While this conjecture implies both the happy edge conjecture~\cite{AichholzerBBDDKLLTU24} and the slightly weaker reparking conjecture~\cite{structural-properties-GD26}, it was recently disproved by Aichholzer, Dorfer, Kramer, Rieck, and Vogtenhuber alongside the latter~\cite{structural-properties-GD26}.
    Their counterexamples show that shortest flip sequences may insert additional edges that are not edges of the convex hull, and may require an edge to be flipped more than twice, refuting the \emph{reparking conjecture}.
    This invalidated the most recent and promising approach to the general happy edge conjecture.
    For some settings with restricted flip operations, the conjecture has been settled independently.
    In~particular, this is the case for flips restricted to rotations and edge slides~\cite{AichholzerBBDDKLLTU24,AichholzerDV25,aichholzer2007quadratic}.

    This left three cases unresolved: compatible flips on point sets in general position, and unrestricted flips on point sets in either convex or general position.

    From a computational complexity perspective, Bjerkevik, Dorfer, Kleist, Ueckerdt, and Vogtenhuber~\cite{bjerkevik2026flippingHardness} proved that, even on convex point sets, deciding whether two plane spanning trees can be transformed into one another using at most $\ell$ flips is \NP-complete.
    This holds for unrestricted flips, compatible flips, and rotations.

    \subsection{Our contribution}\label{subsec:contribution}
    We resolve the three remaining cases of the happy edge property for plane spanning trees: compatible flips on point sets in general position, and unrestricted flips on point sets in either convex or general position.
    In particular, we give counterexamples in all three settings.

    \medskip
    The first construction consists of two plane spanning trees on $16$ points in general position for which preserving all happy edges increases the flip distance from $12$ to $13$.
    \begin{restatable}{theorem}{thmGeneralCounterexample}\label{thm:generalCounterexample}
        There exists a set $P$ of $16$ points in general position, and two plane spanning trees $T_I$ and $T_F$ on $P$, such that the flip distance between $T_I$ and $T_F$ is $12$, but every flip sequence from $T_I$ and $T_F$ that never removes a happy edge has at least $13$ flips.

        In particular, every shortest flip sequence from $T_I$ and $T_F$ removes the same happy edge.
    \end{restatable}

    Interestingly, both distances are attained by compatible flip sequences, and the lower bounds apply to unrestricted flips.
    Thus the same distances hold when only compatible flips are allowed, giving the following consequence.

    \begin{restatable}{corollary}{corGeneralCounterexampleCompatible}\label{cor:generalCompatibleCounterexample}
        The happy edge property fails for compatible flips of plane spanning trees on point sets in general position.
    \end{restatable}

    The second construction consists of two plane spanning trees on $30$ points in convex position.
    Here, preserving all happy edges increases the flip distance from $19$ to $20$.

    \begin{restatable}{theorem}{thmConvexCounterexample}\label{thm:convexCounterexample}
        There exist two plane spanning trees $T_I$ and $T_F$ on $30$ points in convex position,
        such that the flip distance between $T_I$ and $T_F$ is $19$, but every flip sequence from $T_I$ and $T_F$ that never removes a happy edge has at least $20$ flips.

        In particular, every shortest flip sequence from $T_I$ and $T_F$ removes the same happy edge.
    \end{restatable}

    Both constructions exploit the same principle: temporarily removing a happy edge allows a single parking edge to help transform two different parts of the trees.
    The resulting reduction in the number of parking edges outweighs the additional insertion needed to restore the happy edge.
    We give explicit flip sequences and direct combinatorial proofs of the matching lower bounds.
    Together with the known results for rotations and slides, our results complete the classification for the four prominent flip types.
    The happy edge property holds only for compatible flips on convex point sets, as summarized in~\cref{tab:HEP-spanning-trees-flip-types}.

    \begin{table}[htb]
        \caption{The happy edge property for plane spanning trees
        under the four prominent flip types.
        A check mark indicates that the property holds,
            and a cross indicates that it fails.}
        \centering
        \begin{tabular}{|l|c|c|}
            \hline {{Happy edge property}}
            & {{Convex position}} & {{General position}} \\
            \hline\hline
            {{Flips}} &
            \no~[\cref{thm:convexCounterexample}] &
            \no~[\cref{thm:generalCounterexample}] \\
            \hline
            {{Compatible flips}} &
            \yes~\cite{AichholzerDV25} &
            \no~[\Cref{cor:generalCompatibleCounterexample}] \\
            \hline
            {{Rotations}} &
            \no~\cite{AichholzerDV25} &
            \no~\cite{AichholzerDV25} \\
            \hline
            {{Slides}} &
            \no~\cite{AichholzerBBDDKLLTU24} &
            \no~\cite{aichholzer2007quadratic} \\
            \hline
        \end{tabular}
        \label{tab:HEP-spanning-trees-flip-types}
    \end{table}

    \subsection{Definitions and concepts}\label{subsec:definitions}

    We consider the reconfiguration of plane spanning trees on general point sets and point sets in convex position.
    Let $P$ denote a planar point set.
    For $u,v\in P$, we denote by~$uv$ the straight-line segment connecting the two points.
    A \emph{plane spanning tree} of $P$ is then a set of~${n-1}$ edges $T\subset {P\choose 2}$ that define a tree in which the edges' line segments are pairwise either disjoint or meet at a common endpoint.
    We denote the removal and insertion of an edge $e$ by $T-e$ and $T+e$, using standard set notation otherwise.
    A \emph{flip} in a plane spanning tree $T$ removes an edge $e\in T$ and adds an edge $f\notin T$ such that $T-e+f$ is again a plane spanning tree.
    If $T$ is clear from context, we denote this by $e\rightarrow f$ and write that \emph{$e$ is flipped to $f$}.
    For a graph~$G$, we denote its vertex set by $V(G)$ and its edge set by $E(G)$.

    \begin{figure}[htb]%
        \centering%
        \includegraphics[page=1]{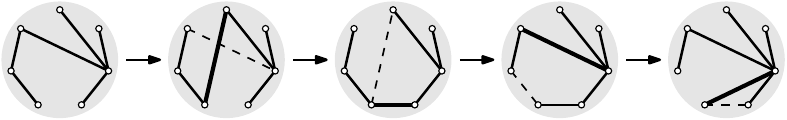}%
        \caption{%
            Flip sequence \flipSeq with four flip types: crossing, rotation, compatible, and edge slide.
        }
        \label{fig:intro-flips}%
    \end{figure}

    \subparagraph*{Flip types.}
    A flip~$e\rightarrow f$ is \emph{compatible} if the two straight-line segments do not share an interior point, and \emph{crossing} otherwise.
    For $u,v,w\in P$, a compatible flip $uv\rightarrow uw$ is a \emph{rotation} {around the vertex $u$.} Furthermore {a flip is} called a \emph{slide} if $vw\in T$.
    This can be interpreted as ``sliding'' the edge $uv$ along $vw$ into $uw$.
    All flip types are depicted in~\cref{fig:intro-flips}.

    \subparagraph*{Flip graphs.}
    The \emph{flip graph} $\mathcal{F}(P)$ of plane spanning trees of $P$ is the graph that has all plane spanning trees as its vertices.
    Two spanning trees share an edge in the flip graph if and only if they can be transformed into one another via a single flip (of some type).
    Given a pair of trees~$T_I$ and~$T_F$, a~\emph{flip sequence}~$\flipSeq$ from~$T_I$ to $T_F$ is then a path in the flip graph denoted by a sequence of trees~${T_I=T_0,T_1,\ldots,T_{k-1},T_k=T_F}$ such that consecutive trees only differ by a single flip; its length is the number $k$ of flips in the sequence.
    The \emph{flip distance} of~$T_I$ and $T_F$, denoted by $d(T_I,T_F)$, is the length of a shortest flip sequence from~$T_I$ to $T_F$.
    The~\emph{diameter} of a flip graph denotes the maximum flip distance between any pair of trees.
    
    An edge~$e\in T_I\cap T_F$ is called a \emph{happy edge} with respect to $T_I$ and $T_F$.
    Conversely, if~$e$ is introduced by a flip, but is not in $T_I\cup T_F$, we say that $e$ is a \emph{parking~edge}.
    For an edge~$e$, we denote by~$\mathcal{F}_{e}(P) := \mathcal{F}(P)\bigl[\{T \in V(\mathcal{F}(P)) \mid e \in T\}\bigr]$ be the subgraph induced by the trees containing $e$.
    For two trees $T_I$ and $T_F$ containing $e$, we denote by $d_{e}(T_I,T_F) := d_{\mathcal{F}_{e}(P)}(T_I,T_F)$ the minimum length of a flip sequence from $T_I$ to $T_F$ that preserves $e$.

    \subsection{Further related work}\label{subsec:relwork}

    Happy edges have a long history in the reconfiguration of plane spanning graphs and for a long time, it seemed like the happy edge property has a strong connection to the computational complexity of finding shortest flip sequences between two configurations.
    Sleator, Tarjan, and Thurston~\cite{SleatorTT86} established the happy edge property for triangulations of convex point sets.
    They also showed that whenever a flip inserts an edge of the target triangulation, some shortest flip sequence starts with that flip.
    Thus, target edges can be inserted greedily whenever possible, without increasing the number of flips needed.
    These insights underpin several \FPT\ algorithms~\cite{bosch2023improved,DBLP:journals/ipl/ClearyJ09,li20253,lucas2010improved}, parameterized by the flip distance.
    Conversely, for triangulations of point sets in general position, the happy edge property fails~\cite{bose2009flipsinplanar}.
    Constructions exploiting this were key to the \NP-hardness proofs in the case of point sets in general position~\cite{lubiw2015flip,pilz2014flip}, as well as simple polygons~\cite{aichholzer2015flip}.

    A similar pattern appears for plane perfect matchings.
    On point sets in convex position, the happy edge property holds, and Hernando, Hurtado, and Noy~\cite{hernando2002graphs} gave a characterization of the flip distance between matchings that allows it to be computed in polynomial time.
    Conversely, for point sets in general position, the happy edge property fails and computing the flip distance is \NP-hard~\cite{binucci2025flipping}.
    For plane almost-perfect matchings, the happy edge property also fails, even in convex position, and deciding whether the flip distance is at most a given bound is \NP-complete on point sets in general position~\cite{aichholzer2025flipping}.

    The aforementioned results suggested that preserving happy edges might be closely linked to the existence of efficient algorithms for computing flip distances.
    Recent results show that this connection does not hold in either direction.
    Aichholzer and Dorfer~\cite{aichholzer2025lineartimealgorithmfinding} gave a linear-time algorithm for computing the flip distance between plane spanning paths on convex point sets, although the happy edge property already fails on four points.
    Conversely, Dorfer~\cite{dorfer2026flipdistancetriangulationsconvex} proved that computing the flip distance between triangulations of convex point sets is \NP-hard, despite the happy edge property.
    The same combination occurs for compatible flips of plane spanning trees on convex point sets: the happy edge property holds, but computing the flip distance is \NP-hard~\cite{bjerkevik2026flippingHardness}.
    Thus the happy edge property alone does not determine the computational complexity of finding shortest flip sequences.
    \cref{tab:HEP-plane-graphs} summarizes the results.

    \begin{table}[htb]
        \caption{Happy edge properties of flips in other families of plane spanning graphs.}
        \label{tab:HEP-plane-graphs}
        \centering
        \begin{tabular}{|l|c|c|}
            \hline {{Happy edge property}}
            & {{Convex position}} & {{General position}} \\
            \hline\hline
            {{Triangulations}} &
            \yes~\cite{SleatorTT86} &
            \no~\cite{bose2009flipsinplanar,hurtado1996flipping} \\
            \hline
            {{Perfect matchings}} &
            \yes~\cite{hernando2002graphs} &
            \no~\cite{schuessler-personal} \\
            \hline
            {{Almost perfect matchings}} &
            \no~\cite{aichholzer2025flipping} &
            \no~\cite{aichholzer2025flipping} \\
            \hline
            {{Paths}} &
            \no~\cite{aichholzer2025lineartimealgorithmfinding} &
            \no~\cite{felsner-personal} \\
            \hline
        \end{tabular}
    \end{table}

    \section{Counterexample for general point sets}\label{sec:general}

    In this section, we provide a counterexample for the happy edge property in general point~sets.
    We present the instance on $16$ points and give an intuitive explanation of why happy edges must be flipped in any shortest flip sequence.
    Afterward, we show that, for this instance, a shortest flip sequence preserving all happy edges requires exactly one more flip than a shortest flip sequence that removes at least one happy edge.

\begin{figure}[htb]%
        \captionsetup[subfigure]{justification=centering}%
        \begin{subfigure}[t]{0.5\columnwidth - 0.5em}%
            \centering%
            \includegraphics[page=1]{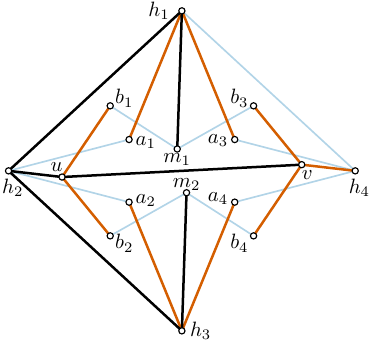}
            \label{fig:generalCounterexampleOverview-a}
            \subcaption{}
        \end{subfigure}%
        \hfill%
        \begin{subfigure}[t]{0.5\columnwidth - 0.5em}%
            \centering%
            \includegraphics[page=2]{figures/counterexample-general}
            \label{fig:generalCounterexampleOverview-b}
            \subcaption{}
        \end{subfigure}
        \caption{The two plane spanning trees $T_I$ (a) and $T_F$ (b) on a point set of $16$ points that constitute a counterexample to the happy edge property in general point sets.
        The happy edge $uv$ must be flipped in every shortest flip sequence.
        We use the following color convention: edges in~$T_I\setminus T_F$ are red, those in $T_F\setminus T_I$ are blue, their intersection is black, and all other edges are green.}
        \label{fig:generalCounterexampleOverview}
    \end{figure}

    \subparagraph*{Construction of the counterexample.}
    The instance consists of two plane spanning trees on a set of $16$ points in general position, both trees are depicted in~\cref{fig:generalCounterexampleOverview}.
    Six of the $15$ edges are happy edges.
    Thus, every flip sequence transforming $T_I$ into $T_F$ requires at least nine flips as every flip can only insert one edge of $T_F\setminus T_I$.

    \subparagraph*{High-level intuition.}
    The construction contains four pairs $\{a_i,b_i\}$ of leaves whose initial and target edges mutually block one another.
    Thus, each pair needs a parking edge to maintain connectivity while the obstruction is resolved.
    Every edge connecting vertices from different pairs crosses a happy edge, so preserving all happy edges forces the four pairs to use distinct parking edges.
    Removing the happy edge $uv$ allows two parking edges to serve two pairs of leaves each, first keeping one leaf connected, and later another one.
    Sharing those parking edges saves two insertions, while restoring the happy edge $uv$ only costs one.
    Together, this yields a sequence of~${9+2+1 = 12}$ flips.
    Since preserving all happy edges results in at least~${9+4 = 13}$ flips, every shortest flip sequence must remove at least one happy edge.

    \subparagraph*{Why preserving happy edges requires 13 flips.}
    We first show that each pair $\{a_i,b_i\}$ of leaves requires an incident parking edge.
    We then use this observation to prove that every flip sequence preserving all happy edges has length at least $13$.

    For each $p\in\{a_i,b_i\},\ 1\le i\le4$, let $e_I(p)$ and $e_F(p)$ denote its unique incident edges in~$T_I$ and $T_F$, respectively.
    Recall that, for each $i$, $e_F(a_i)$ crosses $e_I(b_i)$, and $e_F(b_i)$ crosses~$e_I(a_i)$, as visualized in~\cref{fig:generalCounterexampleOverview}.

    \begin{lemma}\label{lem:generalParkingEdges}
        Every flip sequence from $T_I$ to $T_F$ contains, for each $i\in\{1,2,3,4\}$, a parking edge incident to at least one vertex of the pair $\{a_i,b_i\}$ of leaves.
    \end{lemma}
    \begin{proof}
        Fix any $i\in\{1,2,3,4\}$ and consider the first flip that removes either $e_I(a_i)$ or $e_I(b_i)$.
        Such a flip exists because neither edge belongs to $T_F$.
        Suppose that it removes $e_I(a_i)$; the other case is symmetric.
        Immediately after the flip, $e_I(b_i)$ is still present, and hence, $e_F(a_i)$ is still missing as well, as it crosses $e_I(b_i)$.
        Since every flip must result in a tree, there is an edge $p$ incident to $a_i$.
        The only edges incident to $a_i$ in $T_I\cup T_F$ are $e_I(a_i)$ and $e_F(a_i)$, but neither of them can be added in that flip.
        Thus, $p$ must be a parking edge.
    \end{proof}

    \begin{lemma}\label{lem:generalLowerBoundHappyEdges}
        Every flip sequence from $T_I$ to $T_F$ that preserves all happy edges has length at least $13$.
    \end{lemma}

    \begin{proof}
        Consider a flip sequence $\flipSeq$ that preserves all happy edges.
        By~\cref{lem:generalParkingEdges}, for each~${i\in\{1,2,3,4\}}$, the sequence contains a parking edge $p_i$ incident to at least one vertex of the pair $\{a_i,b_i\}$.
        Every edge connecting vertices from different pairs crosses a happy edge.
        Consequently, the parking edges $p_i$ for $i\in\{1,2,3,4\}$ are distinct.
        Each of those four parking edges must be inserted during the flip sequence, in addition to the nine edges that are in $T_F$ but not in $T_I$.
        Since a flip can only insert a single edge, the flip sequence $\flipSeq$ has length at least $9+4=13$.
    \end{proof}

\begin{figure}[p]%
        \captionsetup[subfigure]{justification=centering}%
        \begin{subfigure}[t]{\textwidth/4}%
            \centering%
            \includegraphics[page=1]{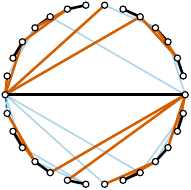}%
            \caption{$T_I\to T_1$}
        \end{subfigure}%
        \begin{subfigure}[t]{\textwidth/4}%
            \centering%
            \includegraphics[page=2]{counterexample-convex-sequence}%
            \subcaption{$T_1\to T_2$}
        \end{subfigure}%
        \begin{subfigure}[t]{\textwidth/4}%
            \centering%
            \includegraphics[page=3]{counterexample-convex-sequence}%
            \subcaption{$T_2\to T_3$}
        \end{subfigure}%
        \begin{subfigure}[t]{\textwidth/4}%
            \centering%
            \includegraphics[page=4]{counterexample-convex-sequence}%
            \subcaption{$T_3\to T_4$}
        \end{subfigure}%
        \par\vspace{1.5em}%
        \begin{subfigure}[t]{\textwidth/4}%
            \centering%
            \includegraphics[page=5]{counterexample-convex-sequence}%
            \caption{$T_4\to T_5$}
        \end{subfigure}%
        \begin{subfigure}[t]{\textwidth/4}%
            \centering%
            \includegraphics[page=6]{counterexample-convex-sequence}%
            \subcaption{$T_5\to T_6$}
        \end{subfigure}%
        \begin{subfigure}[t]{\textwidth/4}%
            \centering%
            \includegraphics[page=7]{counterexample-convex-sequence}%
            \subcaption{$T_6\to T_7$}
        \end{subfigure}%
        \begin{subfigure}[t]{\textwidth/4}%
            \centering%
            \includegraphics[page=8]{counterexample-convex-sequence}%
            \subcaption{$T_7\to T_8$}
        \end{subfigure}%
        \par\vspace{1.5em}%
        \begin{subfigure}[t]{\textwidth/4}%
            \centering%
            \includegraphics[page=9]{counterexample-convex-sequence}%
            \caption{$T_8\to T_9$}
        \end{subfigure}%
        \begin{subfigure}[t]{\textwidth/4}%
            \centering%
            \includegraphics[page=10]{counterexample-convex-sequence}%
            \subcaption{$T_{9}\to T_{10}$}
        \end{subfigure}%
        \begin{subfigure}[t]{\textwidth/4}%
            \centering%
            \includegraphics[page=11]{counterexample-convex-sequence}%
            \subcaption{$T_{10}\to T_{11}$}
        \end{subfigure}%
        \begin{subfigure}[t]{\textwidth/4}%
            \centering%
            \includegraphics[page=12]{counterexample-convex-sequence}%
            \subcaption{$T_{11}\to T_{12}$}
        \end{subfigure}%
        \par\vspace{1.5em}%
        \begin{subfigure}[t]{\textwidth/4}%
            \centering%
            \includegraphics[page=13]{counterexample-convex-sequence}%
            \caption{$T_{12}\to T_{13}$}
        \end{subfigure}%
        \begin{subfigure}[t]{\textwidth/4}%
            \centering%
            \includegraphics[page=14]{counterexample-convex-sequence}%
            \subcaption{$T_{13}\to T_{14}$}
        \end{subfigure}%
        \begin{subfigure}[t]{\textwidth/4}%
            \centering%
            \includegraphics[page=15]{counterexample-convex-sequence}%
            \subcaption{$T_{14}\to T_{15}$}
        \end{subfigure}%
        \begin{subfigure}[t]{\textwidth/4}%
            \centering%
            \includegraphics[page=16]{counterexample-convex-sequence}%
            \subcaption{$T_{15}\to T_{16}$}
        \end{subfigure}%
        \par\vspace{1.5em}%
        \hfil%
        \begin{subfigure}[t]{\textwidth/4}%
            \centering%
            \includegraphics[page=17]{counterexample-convex-sequence}%
            \caption{$T_{16}\to T_{17}$}
        \end{subfigure}%
        \begin{subfigure}[t]{\textwidth/4}%
            \centering%
            \includegraphics[page=18]{counterexample-convex-sequence}%
            \subcaption{$T_{17}\to T_{18}$}
        \end{subfigure}%
        \begin{subfigure}[t]{\textwidth/4}%
            \centering%
            \includegraphics[page=19]{counterexample-convex-sequence}%
            \subcaption{$T_{18}\to T_{F}$}
        \end{subfigure}%
        \caption{A shortest flip sequence from $T_I$ to $T_F$.}
        \label{fig:convex-counterexample-opt}
    \end{figure}

    \subparagraph*{A flip sequence of length 12.}
    Due to~\cref{lem:generalLowerBoundHappyEdges}, there are no flip sequences of length less than $13$ that preserve all happy edges.
    However, we now show that there exists a flip sequence of length at most~$12$ that successfully reconfigures $T_I$ into $T_F$.
    Consequently, this flip sequence removes a happy edge.
    This flip sequence, which removes the happy edge $uv$, is listed in~\cref{tab:generalCounterexampleShortSequence} and illustrated in full on~\cref{fig:convex-counterexample-opt}.
    Together with the bound from~\cref{lem:generalLowerBoundHappyEdges}, this already proves that~\cref{conj:happy} is false for plane spanning trees on general point sets.
    In the following, we show that there is in fact no shorter flip sequence, and that the edge $uv$ is the unique happy edge that must be flipped in every shortest flip sequence.

    \begin{table}[htb]%
        \centering
        \caption{
            A sequence of $12$ compatible flips transforming $T_I$ into $T_F$.
            After removing $uv$, we transform pairs $\{a_1,b_1\}$ and $\{a_2,b_2\}$ using the parking edge $a_1a_2$, and afterward the pairs $\{a_3,b_3\}$ and $\{a_4,b_4\}$ using the parking edge $a_3a_4$.
            The final flip restores $uv$.
        }
        \begin{tabular}{|c|ccc|}
            \hline
            \cellcolor{gray!20}{No.} & \cellcolor{gray!20}{removed edge}  & \cellcolor{gray!20} & \cellcolor{gray!20}{inserted edge} \\\hline\hline
            1                                   & $uv$                                       & $\rightarrow$       & $h_4h_1$                                    \\\hline
            2                                   & $e_I(a_1)$                                 & $\rightarrow$       & $a_1a_2$                                    \\\hline
            3                                   & $e_I(b_1)$                                 & $\rightarrow$       & $e_F(b_1)$                                  \\\hline
            4                                   & $e_I(a_2)$                                 & $\rightarrow$       & $e_F(a_1)$                                  \\\hline
            5                                   & $e_I(b_2)$                                 & $\rightarrow$       & $e_F(b_2)$                                  \\\hline
            6                                   & $a_1a_2$                                   & $\rightarrow$       & $e_F(a_2)$                                  \\\hline
            7                                   & $e_I(a_3)$                                 & $\rightarrow$       & $a_3a_4$                                    \\\hline
            8                                   & $e_I(b_3)$                                 & $\rightarrow$       & $e_F(b_3)$                                  \\\hline
            9                                   & $e_I(a_4)$                                 & $\rightarrow$       & $e_F(a_3)$                                  \\\hline
            10                                  & $e_I(b_4)$                                 & $\rightarrow$       & $e_F(b_4)$                                  \\\hline
            11                                  & $a_3a_4$                                   & $\rightarrow$       & $e_F(a_4)$                                  \\\hline
            12                                  & $vh_4$                                     & $\rightarrow$       & $uv$                                        \\\hline
        \end{tabular}
        \label{tab:generalCounterexampleShortSequence}
    \end{table}

    \begin{figure}[p]%
        \captionsetup[subfigure]{justification=centering}%
        \begin{subfigure}[t]{\textwidth/3}%
            \centering%
            \includegraphics[page=3]{counterexample-general}%
            \caption{$T_I\to T_1$}
        \end{subfigure}%
        \begin{subfigure}[t]{\textwidth/3}%
            \centering%
            \includegraphics[page=4]{counterexample-general}%
            \subcaption{$T_1\to T_2$}
        \end{subfigure}%
        \begin{subfigure}[t]{\textwidth/3}%
            \centering%
            \includegraphics[page=5]{counterexample-general}%
            \subcaption{$T_2\to T_3$}
        \end{subfigure}%
        \par\vspace{1.5em}%
        \begin{subfigure}[t]{\textwidth/3}%
            \centering%
            \includegraphics[page=6]{counterexample-general}%
            \subcaption{$T_3\to T_4$}
        \end{subfigure}%
        \begin{subfigure}[t]{\textwidth/3}%
            \centering%
            \includegraphics[page=7]{counterexample-general}%
            \subcaption{$T_4\to T_5$}
        \end{subfigure}%
        \begin{subfigure}[t]{\textwidth/3}%
            \centering%
            \includegraphics[page=8]{counterexample-general}%
            \subcaption{$T_5\to T_6$}
        \end{subfigure}%
        \par\vspace{1.5em}%
        \begin{subfigure}[t]{\textwidth/3}%
            \centering%
            \includegraphics[page=9]{counterexample-general}%
            \subcaption{$T_6\to T_7$}
        \end{subfigure}%
        \begin{subfigure}[t]{\textwidth/3}%
            \centering%
            \includegraphics[page=10]{counterexample-general}%
            \subcaption{$T_7\to T_8$}
        \end{subfigure}%
        \begin{subfigure}[t]{\textwidth/3}%
            \centering%
            \includegraphics[page=11]{counterexample-general}%
            \subcaption{$T_8\to T_9$}
        \end{subfigure}%
        \par\vspace{1.5em}%
        \begin{subfigure}[t]{\textwidth/3}%
            \centering%
            \includegraphics[page=12]{counterexample-general}%
            \subcaption{$T_9\to T_{10}$}
        \end{subfigure}%
        \begin{subfigure}[t]{\textwidth/3}%
            \centering%
            \includegraphics[page=13]{counterexample-general}%
            \subcaption{$T_{10}\to T_{11}$}
        \end{subfigure}%
        \begin{subfigure}[t]{\textwidth/3}%
            \centering%
            \includegraphics[page=14]{counterexample-general}%
            \subcaption{$T_{11}\to T_F$}
        \end{subfigure}%
        \caption{A shortest flip sequence from $T_I$ to $T_F$. Note that the first flip removes the happy edge~$uv$, which is then reinserted by the final flip of the sequence.}
        \label{fig:general-counterexample-opt}
    \end{figure}

    \thmGeneralCounterexample*
    \begin{proof}
        The proof is split into three parts.
        We first show that (1) the flip distance is $12$.
        We~then prove that (2) preserving all happy edges requires exactly $13$ flips.
        Finally, we show that (3) every shortest flip sequence removes $uv$ and preserves all other happy edges.

        By~\cref{lem:generalParkingEdges}, every flip sequence contains a parking edge incident to a vertex of each of the four pairs $\{a_i,b_i\}$ of leaves.
        Since an edge connects two vertices, every sequence contains at least two distinct parking edges.

        We begin by showing that $12$ flips are necessary.
        A sequence preserving all happy edges has length at least $13$ by~\cref{lem:generalLowerBoundHappyEdges}.
        Any other sequence must reinsert a happy edge that it removes, since it belongs to the target tree.
        Additionally, the sequence must also insert the nine edges that belong to $T_F$ but are not in $T_I$, and at least two distinct parking edges.
        It~therefore has length at least $1+9+2=12$.
        Together with the sequence given in~\cref{tab:generalCounterexampleShortSequence}, this proves that $d(T_I,T_F) = 12$.

        We now prove that $13$ flips suffice when all happy edges must be preserved.
        The first flip~${vh_4 \rightarrow h_1h_4}$ inserts a target edge.
        Afterward, we reconfigure the four pairs of leaves, one after the other, using three flips for each pair.
        For $i\in\{1,2,3,4\}$, let
        \[
            p_i=
            \begin{cases}
                a_i u, & i\in\{1,2\},\\
                a_i v, & i\in\{3,4\}.
            \end{cases}
        \]
        Thus $p_i$ joins $a_i$ to the initial neighbor of $b_i$.
        For each $i$, in increasing order, perform the flips
        \[
            e_I(a_i)\rightarrow p_i,\quad
            e_I(b_i)\rightarrow e_F(b_i),\quad
            p_i\rightarrow e_F(a_i).
        \]
        The first flip keeps $a_i$ connected through the parking edge $p_i$ while removing $e_I(a_i)$, which crosses $e_F(b_i)$.
        The second flip inserts $e_F(b_i)$ and removes $e_I(b_i)$, which crosses $e_F(a_i)$.
        The~third flip therefore completes the transformation of the pair by replacing $p_i$ with $e_F(a_i)$.
        In every flip, the removed and inserted edges are incident to the same leaf, so the resulting graph remains a spanning tree and the two exchanged edges do not cross.
        In each flip, the inserted and removed edges do not cross.
        Thus all flips are valid and compatible.
        The sequence did not remove any happy edge, and it has length $1+4\cdot3=13$.
        Together with~\cref{lem:generalLowerBoundHappyEdges}, this proves that $d_H(T_I,T_F) = 13$.

        It remains to show that every shortest flip sequence removes $uv$ and preserves every other happy edge.
        For this, consider any flip sequence of length $12$.
        It must remove a happy edge, since otherwise it has length at least $13$.
        We already know that the sequence removes exactly one happy edge.
        Suppose that $uv$ is preserved.
        Neither parking edge can connect a vertex of the pairs $\{a_1,b_1\}$ or $\{a_3,b_3\}$ to a vertex of the pairs $\{a_2,b_2\}$ or $\{a_4,b_4\}$, since such an edge would cross $uv$.
        Nevertheless, the two parking edges must together be incident to all four pairs by~\cref{lem:generalParkingEdges}.
        Consequently, one must join $\{a_1,b_1\}$ and $\{a_3,b_3\}$, and another one joins~$\{a_2,b_2\}$ and $\{a_4,b_4\}$.
        The former crosses $m_1h_1$, the latter crosses $m_2h_3$.
        Both of these happy edges must be removed, contradicting the fact that exactly one happy edge is removed in any flip sequence of length $12$.
        Hence, every shortest flip sequence removes $uv$ and preserves every other happy edge.
    \end{proof}

    Note that all flips in our sequence of $12$ flips are compatible, and that the arguments in the proof of~\cref{thm:generalCounterexample} apply to compatible flips as well.
    Thus, we obtain the~following.
    \corGeneralCounterexampleCompatible*

    \section{Counterexample for convex point sets}\label{sec:convex}
    We now turn our attention to convex point sets.
    Recall that, unlike general points, the happy edge property holds for compatible flips on convex point sets~\cite{AichholzerDV25}.
    We will construct an instance consisting of two trees $T_I$ and $T_F$ on the vertices of a regular $30$-gon, and again give an intuitive explanation of why happy edges must be flipped in any shortest flip sequence.

\begin{figure}[htb]
        \captionsetup[subfigure]{justification=centering}%
        \begin{subfigure}[t]{0.5\textwidth - 0.5em}%
            \centering%
            \includegraphics[page=1]{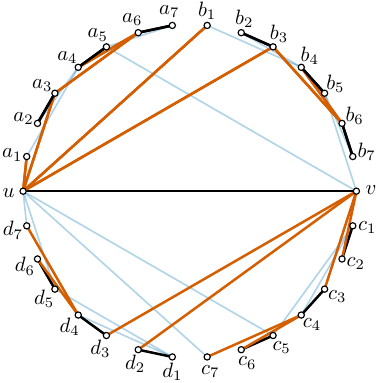}%
            \subcaption{}%
        \end{subfigure}%
        \hfill%
        \begin{subfigure}[t]{0.5\textwidth - 0.5em}%
            \centering%
            \includegraphics[page=2]{figures/counterexample-convex.pdf}%
            \subcaption{}%
        \end{subfigure}%
        \caption{The two plane spanning trees $T_I$ (a) and $T_F$ (b) on a convex point set of $30$ points that constitute a counterexample to the happy edge property.
        The happy chord $uv$ must be flipped in any shortest flip sequence.}
        \label{fig:convexCounterexampleOverview}
    \end{figure}

    \subparagraph*{Construction of the counterexample.}
    Consider $30$ points in convex position in counterclockwise cyclic order.
    We label these~$u,a_1,\dots,a_7,b_1,\dots,b_7,v,c_1,\dots,c_7,d_1,\dots,d_7$, as depicted in~\cref{fig:convexCounterexampleOverview}.
    The happy chord~$uv$ separates the four groups of seven vertices each into two halves, each half containing two of the groups together with $u$ and $v$.
    In $T_I$, the $a$- and $b$-groups are connected to $u$, while the $c$- and $d$-groups attach to $v$.
    This attachment is reversed in $T_F$, and the edges within each group are rearranged.
    The two trees share thirteen edges, the chord~$uv$ as well as twelve convex hull edges; thus, we have an immediate lower bound of $16$ on their flip distance.
    In the $a$- and $b$-groups, every missing target edge crosses two initial edges; for the $c$- and $d$-groups, the same applies in reverse.

\pagebreak
    \subparagraph*{High-level intuition.}
    The construction is built around the happy chord $uv$, which separates the $a$- and $b$-groups from the $c$- and $d$-groups.
    As long as $uv$ remains, no edge can connect a vertex strictly on one side to a vertex strictly on the other.
    Consequently, every flip acts within one half, and the lengths of the flip sequences within both halves add up.

    Each half contains two groups, with four target edges in each group; thus, at least eight flips are needed to insert just these edges.
    However, the groups cannot be transformed by using only those $8$ flips: each target edge in the top half crosses two edges from the initial tree, so we must first remove such an edge and use it as a connectivity preserving parking edge.
    For the lower half, the same argument applies when we read the reconfiguration~backwards.

    We claim that each half needs at least $10$ flips.
    To understand this lower bound, consider the upper half of $P$.
    As noted above, we need to insert eight target edges, but none of them can be inserted in the first flip, as each crosses two existing edges.
    A flip sequence of $9$ flips only has room for a single parking flip, which would have to help both groups.
    However, the construction rules this out:
    any parking edge capable of helping both groups would cross an existing edge that must still be present when the parking edge is inserted.
    Thus, at least $10$ flips are needed.
    The same arguments apply for the bottom half.

    Now, removing $uv$ allows a different approach, as we can work on groups from opposite halves together.
    We first connect the $b$- and $d$-groups with a parking edge.
    This edge first helps to keep parts of the $b$-group connected through the $d$-group while we insert edges in the upper half.
    Simultaneously, it keeps parts of the $d$-group connected to the $b$-group when we insert edges in the bottom half.
    Thus, one parking edge provides a temporary connection that is needed for both groups.

    However, this results in a flip sequence of $10$ flips when we restore $uv$ after those transformations, and the same holds for reconfiguring the $a$- and $c$-groups, resulting in~$20$~flips.
    We can save one flip in between:
    rather than putting $uv$ back after transforming the $b$- and $d$-groups, we directly flip to another parking edge that helps to reconfigure the $a$- and $c$-groups.
    We then restore $uv$ only at the very end.

    Altogether, this gives a flip sequence of $19$ flips, whereas preserving $uv$ requires at least~$20$.
    Consequently, every shortest flip sequence must remove this happy edge.

    \subparagraph*{Preserving happy edges yields flip sequences of length at least 20.}
    Let $P_{ab}$ and $P_{cd}$ denote the two halves of the point set, each including $u$ and $v$, and let $T_j[P_{ab}]$ and $T_j[P_{cd}]$ be the corresponding subgraphs for $j\in\{I,F\}$.

    As long as $uv$ is preserved, every flip acts entirely within one half.
    Indeed, connecting any vertex of $P_{ab}\setminus\{u,v\}$ to any vertex of $P_{cd}\setminus\{u,v\}$ would cross $uv$.
    Consequently, every intermediate tree $T$ induces a spanning tree on each half: every vertex must be connected within its own half to either $u$ or $v$, which are themselves joined by $uv$. Removing an edge other than $uv$ from one of these induced trees disconnects it, and inserting an edge in the other half cannot restore its connectivity.
    From this, it follows that
    \[d_{uv}(T_I,T_F) = d_{uv}\bigl(T_I[P_{ab}],T_F[P_{ab}]\bigr) + d_{uv}\bigl(T_I[P_{cd}],T_F[P_{cd}]\bigr).\]

    We now show that, when $uv$ is preserved, exactly $10$ flips are needed in each half to transform the initial tree into the target tree.
    \cref{tab:convexWithFixedChord} gives sequences of $10$ flips each to transform $T_I[P_{ab}]$ into $T_F[P_{ab}]$ and $T_I[P_{cd}]$ into $T_F[P_{cd}]$.

    \begin{table}[htb]
        \centering
        \caption{A sequence of $20$ flips transforming $T_I$ into $T_F$ while preserving all happy edges.
        While the first $10$ flips act on $P_{ab}$, the remaining $10$ act on $P_{cd}$.
        Parking edges are $ua_2$, $ub_2$, $c_4c_5$, and $d_4d_5$.}
        \label{tab:convexWithFixedChord}
        \begin{minipage}[t]{0.48\textwidth}
            \centering
            Flip sequence on $P_{ab}$
            \par\medskip
            \begin{tabular}{|c|ccc|}
                \hline
                \cellcolor{gray!20}{No.} & \cellcolor{gray!20}{removed edge} & \cellcolor{gray!20} & \cellcolor{gray!20}{inserted edge} \\ \hline\hline
                1                                    & $ua_3$                                     & $\rightarrow$       & $ua_2$                                      \\
                2                                    & $a_3a_6$                                   & $\rightarrow$       & $a_2a_4$                                    \\
                3                                    & $ua_2$                                     & $\rightarrow$       & $a_1a_4$                                    \\
                4                                    & $a_4a_6$                                   & $\rightarrow$       & $a_5a_7$                                    \\
                5                                    & $ub_3$                                     & $\rightarrow$       & $ub_2$                                      \\
                6                                    & $b_3b_6$                                   & $\rightarrow$       & $b_2b_4$                                    \\
                7                                    & $ub_2$                                     & $\rightarrow$       & $b_1b_4$                                    \\
                8                                    & $b_4b_6$                                   & $\rightarrow$       & $b_5b_7$                                    \\
                9                                    & $ub_1$                                     & $\rightarrow$       & $b_5v$                                      \\
                10                                   & $ua_1$                                     & $\rightarrow$       & $a_5v$                                      \\ \hline
            \end{tabular}
        \end{minipage}
        \hfill
        \begin{minipage}[t]{0.48\textwidth}
            \centering
            Flip sequence on $P_{cd}$
            \par\medskip
            \begin{tabular}{|c|ccc|}
                \hline
                \cellcolor{gray!20}{No.} & \cellcolor{gray!20}{removed edge} & \cellcolor{gray!20} & \cellcolor{gray!20}{inserted edge} \\ \hline\hline
                11                                   & $vd_3$                                     & $\rightarrow$       & $ud_7$                                      \\
                12                                   & $vd_2$                                     & $\rightarrow$       & $d_1d_4$                                    \\
                13                                   & $vc_3$                                     & $\rightarrow$       & $uc_7$                                      \\
                14                                   & $vc_2$                                     & $\rightarrow$       & $c_1c_4$                                    \\
                15                                   & $c_4c_6$                                   & $\rightarrow$       & $c_4c_5$                                    \\
                16                                   & $c_4c_7$                                   & $\rightarrow$       & $uc_5$                                      \\
                17                                   & $c_4c_5$                                   & $\rightarrow$       & $c_1c_5$                                    \\
                18                                   & $d_4d_6$                                   & $\rightarrow$       & $d_4d_5$                                    \\
                19                                   & $d_4d_7$                                   & $\rightarrow$       & $ud_5$                                      \\
                20                                   & $d_4d_5$                                   & $\rightarrow$       & $d_1d_5$                                    \\ \hline
            \end{tabular}
        \end{minipage}
    \end{table}

    \begin{lemma}\label{lem:independentLongSequences}
        For each of the two halves $P_{ab}$ and $P_{cd}$, exactly~$10$ flips are needed to transform the induced initial tree into the induced target tree while preserving $uv$.
    \end{lemma}
    \begin{proof}
        \cref{tab:convexWithFixedChord} describes sequences of length $10$ for the two halves, each preserving $uv$.
        Thus, it remains to show that fewer than $10$ flips cannot suffice.

        We first consider the reconfiguration from $T_I[P_{ab}]$ to $T_F[P_{ab}]$.
        For each $z\in\{a,b\}$, every target edge in $\{z_1z_4,z_2z_4\}$ crosses both initial edges in~$\{uz_3,z_3z_6\}$, and every target edge in~$\{z_5z_7,z_5v\}$ crosses both initial edges in $\{z_3z_6,z_4z_6\}$.
        Thus none of the eight missing target edges can be inserted in the first flip, since a flip removes only one edge.

        A sequence of at most nine flips must therefore have exactly nine flips:
        the first inserts a parking edge $p$, and the remaining eight insert the eight missing target edges, each exactly once.
        In particular, $p$ is the only parking edge, and no removed initial edge can be reinserted.

        We show that $p$ must connect vertices from the two groups.
        Consider either group immediately before its first missing target edge is inserted.
        At least one of the two initial edges crossing that target edge must already have been removed, since the
        upcoming flip can remove only one of them.
        Initially, the group together with $u$ induces a tree that is attached to the rest of the vertices only through $u$.
        The removal therefore disconnects some of its vertices from $u$ within the remaining initial edges.
        Recall that no missing target edge of this group has yet been inserted, and that target edges of the other group have no endpoint in this group.
        Thus, $p$ must be present and reconnect these vertices.
        Applying this argument to both groups shows that $p$ joins an $a$-vertex to a $b$-vertex.

        We now examine the $b$-group more closely.
        Consider the tree immediately before the first missing target edge which is incident to a $b$ vertex is inserted.
        As argued above, at least one initial edge of the group has already been removed.
        We claim that exactly one has been removed.
        Initially, the vertices $u,b_1,\ldots,b_7$ induce a tree.
        Removing two or more edges from this tree would split it into at least three components, at least two of which do not contain $u$.
        Each of these components would need a connection to the rest of the current spanning tree.
        However, no missing target edge of the $b$-group has yet been inserted, and target edges of the $a$-group have no endpoint in the $b$-group.
        Thus, the parking edge $p$ is the only edge that can provide such a connection.
        Since $p$ has exactly one endpoint in the $b$-group, it can connect only one of these components to the rest of the tree.
        Consequently, exactly one initial edge of the $b$-group has been removed.

        The removed edge must be one of the two edges crossing the target edge about to be inserted; otherwise, both crossing edges would still be present, and the upcoming flip could remove only one of them.
        By the crossing relations identified above, the removed edge therefore belongs to $\{ub_3,b_3b_6,b_4b_6\}$.
        In particular, it is not $ub_1$, so $ub_1$ remains present and connects $b_1$ directly to $u$.
        Removing that edge splits the initial tree on $u,b_1,\ldots,b_7$ into two components: one contains both $u$ and $b_1$, and the other must be connected to the rest of the current tree by the parking edge $p$.
        Thus the endpoint of $p$ in the $b$-group must lie in the latter component and cannot be $b_1$.

        We already know that $p$ connects an $a$-vertex to a $b$-vertex.
        Consequently, $p=a_i b_j$ with $1\le i\le7$ and $2\le j\le7$.
        But the endpoints of $p$ and $ub_1$ occur in the cyclic order $u,a_i,b_1,b_j$, so these two edges cross.
        Since both are present in the current tree, this contradicts planarity. 
        Thus, in the upper half we can not start the flip sequence by inserting a parking edge $p$, and consequently we cannot obtain a sequence of length $9$.

        For the other half, we apply the same argument to the reverse transformation, i.e., from $T_F[P_{cd}]$ to~$T_I[P_{cd}]$.
        Here, initial and target edges refer to this reversed direction.
        Each group together with $u$ again induces an initial tree attached to the rest only through $u$.
        For~each~$z\in\{c,d\}$, every target edge in $\{vz_2,vz_3\}$ crosses both initial edges in $\{z_1z_4,z_1z_5\}$, and every target edge in $\{z_4z_6,z_4z_7\}$ crosses both initial edges in $\{uz_5,z_1z_5\}$.

        As before, a sequence of at most nine flips would use a single parking edge $p$ joining the two groups.
        Immediately before the first target edge of the $c$-group is inserted, exactly one edge of that group of the initial tree has been removed.
        By the crossing relations above, it belongs to $\{c_1c_4,c_1c_5,uc_5\}$. 
        Thus $uc_7$ remains present, and $c_7$ lies in the component containing $u$.
        Since $p$ must reconnect the other component, its endpoint in the $c$-group cannot be $c_7$.
        Hence, we have $p=c_i d_j$ with $1\le i\le6$ and $1\le j\le7$.
        Every such edge crosses the preserved edge $uc_7$, since
        their endpoints occur in the cyclic order $c_i,c_7,d_j,u$.
        This is again a contradiction.

        Reversing a sequence preserves its length and the requirement to preserve $uv$, so the forward flip sequence also requires at least $10$ flips.
    \end{proof}

    We have shown that every flip sequence preserving the happy edge $uv$ has length at least~$20$.
    Since all other happy edges lie on the convex hull and are preserved by every shortest flip sequence, any sequence shorter than $20$ flips must remove $uv$.

    \subparagraph*{A flip sequence of length 19 shares parking edges across the happy chord.}
    We now provide a sequence of $19$ flips by temporarily removing $uv$, allowing parking edges to connect vertices on opposite sides of it.
    One such sequence is given in~\cref{tab:convexCounterexampleShortSequence}.

    \begin{table}[htb]
        \centering
        \caption{A sequence of $19$ flips transforming $T_I$ into $T_F$.
        The diagonal $uv$ is removed in the first flip and restored in the last.
        The parking edges $b_4d_4$ and $a_4c_4$ are used successively to transform the four groups.}
        \label{tab:convexCounterexampleShortSequence}
        \begin{tabular}{|c|ccc|}
            \hline
            \cellcolor{gray!20}{No.} & \cellcolor{gray!20}{removed edge} & \cellcolor{gray!20} & \cellcolor{gray!20}{inserted edge} \\\hline\hline
            1                               & $uv$                                       & $\rightarrow$       & $ud_7$                                      \\\hline
            2                               & $b_3b_6$                                   & $\rightarrow$       & $b_4d_4$                                    \\\hline
            3                               & $ub_3$                                     & $\rightarrow$       & $b_2b_4$                                    \\\hline
            4                               & $b_4b_6$                                   & $\rightarrow$       & $b_5b_7$                                    \\\hline
            5                               & $vd_3$                                     & $\rightarrow$       & $b_5v$                                      \\\hline
            6                               & $vd_2$                                     & $\rightarrow$       & $d_1d_4$                                    \\\hline
            7                               & $d_4d_7$                                   & $\rightarrow$       & $b_1b_4$                                    \\\hline
            8                               & $d_4d_6$                                   & $\rightarrow$       & $ud_5$                                      \\\hline
            9                               & $b_4d_4$                                   & $\rightarrow$       & $d_1d_5$                                    \\\hline
            10                              & $ub_1$                                     & $\rightarrow$       & $uc_7$                                      \\\hline
            11                              & $a_3a_6$                                   & $\rightarrow$       & $a_4c_4$                                    \\\hline
            12                              & $ua_3$                                     & $\rightarrow$       & $a_2a_4$                                    \\\hline
            13                              & $a_4a_6$                                   & $\rightarrow$       & $a_5a_7$                                    \\\hline
            14                              & $vc_3$                                     & $\rightarrow$       & $a_5v$                                      \\\hline
            15                              & $vc_2$                                     & $\rightarrow$       & $c_1c_4$                                    \\\hline
            16                              & $c_4c_7$                                   & $\rightarrow$       & $a_1a_4$                                    \\\hline
            17                              & $c_4c_6$                                   & $\rightarrow$       & $uc_5$                                      \\\hline
            18                              & $a_4c_4$                                   & $\rightarrow$       & $c_1c_5$                                    \\\hline
            19                              & $ua_1$                                     & $\rightarrow$       & $uv$                                        \\\hline
        \end{tabular}
    \end{table}

    We now rigorously prove that every shortest flip sequence between the two trees in the construction must remove a happy edge.
    
    \thmConvexCounterexample*
    \begin{proof}
        We first determine the minimum length of a flip sequence preserving $uv$.
        Every such flip acts entirely within one of the two halves $P_{ab}$ and $P_{cd}$.
        By~\cref{lem:independentLongSequences}, each half requires at least ten flips.
        The sequence in~\cref{tab:convexWithFixedChord} attains the resulting bound of $20$ flips and preserves all happy edges.
        Consequently, $d_{uv}(T_I,T_F)=20$.

        We next construct a sequence of $19$ flips by temporarily removing $uv$.
        \cref{tab:convexCounterexampleShortSequence} gives an explicit sequence from $T_I$ to $T_F$.
        The first flip replaces $uv$ by $ud_7$.
        Flips $2$--$9$ transform the $b$- and $d$-groups using the parking edge $b_4d_4$.
        Flip $10$ replaces $ub_1$ by $uc_7$, after which flips~$11$--$18$ transform the $a$- and $c$-groups using the parking edge $a_4c_4$.
        The final flip replaces $ua_1$ by $uv$.
        It can be checked directly that this sequence is valid, giving $d(T_I,T_F)\le 19$.

        To prove the matching lower bound, we show that every sequence contains a parking edge incident to each of the four groups.
        To see this, consider the $a$- or $b$-group just before its first missing target edge is inserted.
        One of the two initial edges crossing this target edge must already have been removed and not yet reinserted.
        This disconnects some vertices from $u$ within the remaining initial edges. 
        Since no missing target edge of the group has yet been inserted, only a parking edge can reconnect them.
        The same argument applies to the $c$- and $d$-groups when the sequence is reversed; reversal does not change which edges are parking~edges.

        An edge has endpoints in at most two groups, so at least two distinct parking edges must occur.
        A sequence preserving $uv$ requires at least $20$ flips.
        Any other sequence must insert the sixteen missing target edges, at least two parking edges, and $uv$ once more. It therefore requires at least $19$ flips.
        Together with the explicit sequence from~\cref{tab:convexCounterexampleShortSequence}, this gives~$d(T_I,T_F)=19$.
        Hence every shortest flip sequence removes the happy edge $uv$, disproving the happy edge property for plane spanning trees on point sets in convex position.
    \end{proof}

    \section{Conclusion}\label{sec:conclusion}
    We have shown that the happy edge property fails for plane spanning trees even on point sets in convex position.
    For point sets in general position, the demonstrated counterexample also applies when only compatible flips are allowed.
    In contrast, the happy edge property holds for compatible flips on point sets in convex position~\cite{AichholzerDV25}.
    Together with the known positive results and counterexamples to the parking edge and reparking conjectures~\cite{structural-properties-GD26}, these results complete the picture of when these three conjectures hold.

    \subparagraph*{AI disclosure.}
    An initial counterexample to the happy edge conjecture for point sets in general position, consisting of $29$ points, was found using ChatGPT-Sol 6.1.
    The smaller counterexample in general position as well as the counterexample in convex position were found using Claude~Opus~5.5.
    The manuscript and figures were prepared by the authors. 
    We refined, extended, and simplified the reasoning, and take full responsibility for the mathematical correctness of the presented arguments.

    \bibliography{references}
\end{document}